\documentclass[onecolumn]{aastex631}

\usepackage{amsmath,amsthm,amssymb,amsfonts,xspace,graphicx}
\usepackage{fontawesome}

\shorttitle{A point cloud approach to field level generative modeling}
\shortauthors{Cuesta-Lazaro \& Mishra-Sharma}

\reportnum{MIT-CTP/5651}

\newcommand{\package}[1]{\textsl{#1}\xspace}

\newcommand{\nbicon}{{\color{xlinkcolor}\faFileCodeO}\xspace}

\newcommand{\githubmaster}{\href{https://github.com/smsharma/point-cloud-galaxy-diffusion}{\faGithub}\xspace}
\newcommand{\nblink}[1]{\href{https://github.com/smsharma/point-cloud-galaxy-diffusion/blob/arXiv-v1/notebooks/#1.ipynb}{\nbicon}}

\newcommand{\iaifilogo}{\includegraphics[width=0.8em]{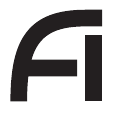}\xspace}

\DeclareMathOperator*{\argmin}{arg\,min} 

\begin{document}\sloppy\sloppypar\raggedbottom\frenchspacing

\title{A point cloud approach to generative modeling for galaxy surveys at the field level}

\author[0000-0002-6069-2999]{Carolina Cuesta-Lazaro}
\altaffiliation{Both authors contributed equally to this work. Authors may be listed in either order.}
\affiliation{\iaifilogo The NSF AI Institute for Artificial Intelligence and Fundamental Interactions}
\affiliation{Department of Physics, Massachusetts Institute of Technology, Cambridge, MA 02139, USA}
\affiliation{Center for Astrophysics | Harvard \& Smithsonian, 60 Garden Street, MS-16, Cambridge, MA 02138, USA}

\author[0000-0001-9088-7845]{Siddharth Mishra-Sharma}
\altaffiliation{Both authors contributed equally to this work. Authors may be listed in either order.}
\affiliation{\iaifilogo The NSF AI Institute for Artificial Intelligence and Fundamental Interactions}
\affiliation{Center for Theoretical Physics, Massachusetts Institute of Technology, Cambridge, MA 02139, USA}
\affiliation{Department of Physics, Massachusetts Institute of Technology, Cambridge, MA 02139, USA}
\affiliation{Department of Physics, Harvard University, Cambridge, MA 02138, USA}

\email{\href{mailto:smsharma@mit.edu}{smsharma@mit.edu} and \href{mailto:cuestalz@mit.edu}{cuestalz@mit.edu}}

\begin{abstract}\noindent
    We introduce a diffusion-based generative model to describe the distribution of galaxies in our Universe directly as a collection of points in 3-D space (coordinates) optionally with associated attributes (e.g., velocities and masses), without resorting to binning or voxelization. The custom diffusion model can be used both for emulation, reproducing essential summary statistics of the galaxy distribution, as well as inference, by computing the conditional likelihood of a galaxy field. We demonstrate a first application to massive dark matter haloes in the \emph{Quijote} simulation suite. This approach can be extended to enable a comprehensive analysis of cosmological data, circumventing limitations inherent to summary statistics- as well as neural simulation-based inference methods. \githubmaster
\end{abstract}

\keywords{
Astrostatistics techniques (1886)
---
Large-scale structure of the universe (902)
}

\section{Introduction}
\label{sec:intro}

Cosmological data analysis is a multidisciplinary field that involves nuanced interplay between theory and data. Analysis of late-time observables of structure formation is especially challenging due to the high dimensionality of typical data and complexity of the underlying data-generating process, which aims to model, amongst others, the nonlinear collapse of structures, baryonic processes, and the formation of galaxies in the dark matter cosmic web. An example of such an observable is galaxy clustering -- the 3-D distribution of galaxies in the Universe -- which is a powerful probe of cosmology and galaxy formation.

The galaxy clustering signal is typically quantified by summary statistics like the two-point correlation function (2PCF), which measures the probability of finding a pair of galaxies as a function of their separation in excess of expectation based on a uniform distribution. While routinely used in cosmological analyses, the 2PCF is not a complete (sufficient) summary of the galaxy clustering signal, and other statistics like higher-order correlation functions \citep{IntroStats_Bispectra_CMBxLSS,IntroStats_PkBiTrispectra}, wavelet scattering transforms \citep{2020MNRAS.499.5902C,IntroStats_Wavelets,Valogiannis:2022xwu}, density statistics \citep{paillas2023cosmological,IntroStats_PDF}, void statistics \citep{IntroStats_Voids_Pisani2019,IntroStats_Voids_Hawken2020}, $k$-nearest neighbour summaries \citep{Banerjee_2020}, and many others are routinely employed to capture additional information contained in the clustering signal, in particular at smaller scales where non-linear structure formation is critical to the description of the field. Recent studies \citep{paillas2023cosmological, valogiannis2023precise} have shown that the information extracted from existing galaxy surveys can be more than doubled through the use of alternative summary statistics that go beyond the 2PCF.

Machine learning methods have demonstrated the potential to significantly impact how cosmological data is analyzed, and galaxy clustering is no exception \citep{Dai_2022, Makinen_2022,Hahn_2023}. More concretely, the ability of neural networks to beat the curse of dimensionality allows for extraction of information about the underlying cosmology without having to manually construct summary statistics to describe the galaxy clustering field.

For galaxy clustering observations, arguably the holy grail is to obtain a reliable likelihood of an observed galaxy configuration $x$ given some parametric description $\theta$ of underlying cosmological models of interest, $p(x\mid\theta)$ which is additionally amenable to sampling -- a \emph{generative model}. Access to the conditional likelihood can be use to sample different field configurations, $x\sim p(x\mid\theta)$, for use in various downstream tasks or as a surrogate model ({emulation}). Additionally, one can use the likelihood to perform {parameter inference} and hypothesis testing using a method of ones choosing. In the context of Bayesian inference, commonly employed in cosmology, the conditional likelihood can be used in conjunction with a prior $p(\theta)$ in order to obtain a estimate of the parameter posterior density, $p(\theta\mid x) = p(x\mid\theta)\cdot p(\theta) / p(x)$.

Unfortunately, computing the conditional likelihood is extremely challenging for most observationally interesting scenarios. This is because it requires marginalizing over an essentially infinite-measure space of latent configurations, denoted $z$, characterizing possible initial conditions and their evolution trajectories towards realizing a given observation $x$; $p(x\mid\theta)  = \int \mathrm dz\,p(x,z\mid\theta)$. For a collection of galaxies or dark matter halos, constructing a generative model involves modeling the joint probability density of the properties (positions, velocities, etc.) of a large number of galaxies, $p\left(\{x_i\}_{i=1}^{N_\mathrm{gal}}\mid\theta\right)$, while simultaneously capturing the dependence on cosmology -- a formidable task.

Machine learning has revolutionized the field of generative modeling, heralding methods that are able to learn complex data distributions such as those of natural images and human-generated text. Much of this success has been enabled through the use of {diffusion models}~\citep{ho2020denoising,song2020score} -- a class of generative models that, colloquially, learn to efficiently denoise a corrupted version of the data. Within the sciences, diffusion models have demonstrated potential across domains, showing impressive performance in modeling the distribution of atomistic systems \cite[e.g.,][]{2022arXiv220317003H}, proteins and biomolecules \citep[e.g.,][]{Ingraham2022.12.01.518682,Watson2022.12.09.519842,Alamdari2023.09.11.556673}, and particle jets \citep[e.g.,][]{2023PhRvD.108c6025M,2023arXiv230305376L,2023arXiv231000049B}, to name a few. Compared to other generative models, diffusion models tend to be more expressive than variational autoencoders, allow for more flexible architecture and training than normalizing flows, and can estimate approximate likelihoods unlike generative adversarial networks, while still producing diverse samples.

\begin{figure*}[!t]
    \includegraphics[page=1, width=0.98\textwidth]{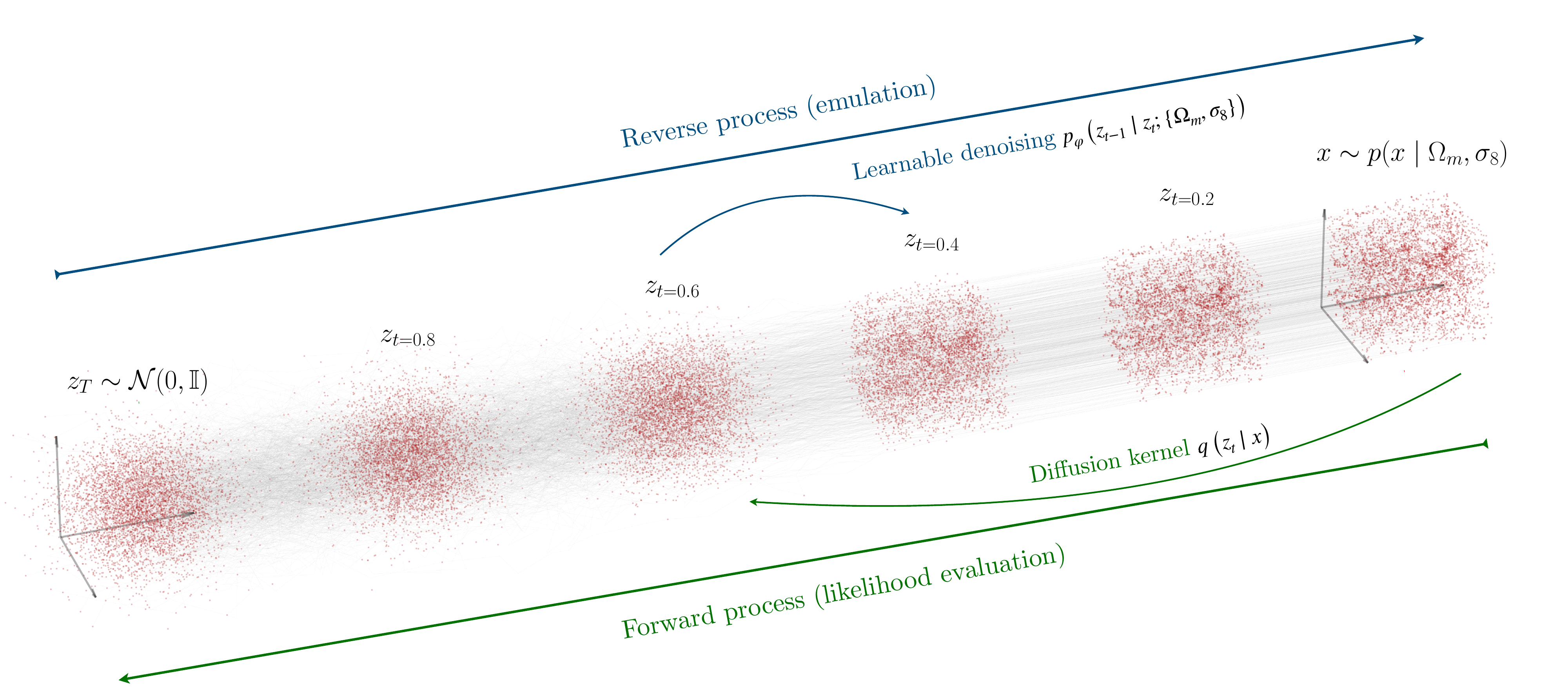}
    \caption{A schematic overview of the point cloud diffusion model, showing samples from the diffusion process at different diffusion times. During training, noise is added to a data sample $x$ using the diffusion kernel $q(z_t\mid x)$ and a denoising distribution $p_\varphi\left(z_{t-1} \mid z_t\right)$ is learned. To generate samples, we simulate the reverse process -- we sample noise from a standard Gaussian distribution and denoise it iteratively using the learned denoising distribution. \nblink{04_viz_pos}}
    \label{fig:samples}
\end{figure*}

Within cosmology, generative modeling has recently been applied in the context of matter density fields~\citep{Dai_2022, mudur2022denoising}, initial conditions reconstruction~\citep{legin2023posterior}, weak lensing mass maps~\citep{remy2022probabilistic}, galaxy images~\citep{lanusse2021deep,smith2022realistic}, and strong lensing observations~\citep{adam2022posterior,legin2023beyond}. In all cases, the common data modality of 2-D or 3-D pixelized images or voxelized boxes is used. While the image representation is appropriate in many cases, such as weak gravitational lensing, the distribution of galaxies is, arguably, ideally represented as a \emph{point cloud} -- a set of points in 3-D space, with additional attributes (e.g., luminosities, velocities, as well as other galaxy properties) attached to them. Pixelization or voxelization necessarily introduces scale cuts, information loss, as well as hyperparameter choices, precluding a full \emph{in-situ} analysis of observed data.

In this paper, we develop a diffusion-generative model with the goal of describing the statistical properties of the distribution of galaxies in our Universe. We focus here on modeling dark matter halos, leaving a more detailed exploration including effects of the galaxy-halo connection and observational effects to future work. We show that our custom diffusion model, which uses either graph neural networks or transformers as a backbone, faithfully reproduces crucial summary properties of the galaxy field with expected cosmological dependence. Furthermore, we show how our model can be used to evaluate the conditional likelihood of a galaxy field. 

This paper is organized as follows. We describe our methodology, including an overview of the diffusion modeling framework, the underlying data-processing neural networks involved, and a description of the dataset and training procedure in Sec. \ref{sec:methodology}. We showcase generated samples and validate their properties in Sec. \ref{sec:experiments}. We describe in detail the methodological limitations of our model and discuss future avenues for improvement in Sec. \ref{sec:prospects}. We conclude in Sec. \ref{sec:conclusions}.

\section{Methodology}
\label{sec:methodology}

We describe, in turn, the simulation dataset used, the diffusion model framework, and the noise-prediction neural network backbones of our model.

\subsection{Dataset and forward model}

Our dataset is derived from the standard latin hypercube set of the \emph{Quijote} suite of $2000$ $N$-body simulations \citep{Quijote_sims} at redshift zero. These simulations follow the evolution of $512^3$ cold dark matter particles in a volume of $\left(1\, h^{-1}\,\mathrm{Gpc}\right)^3$ with periodic boundary conditions from $z=127$ to $z=0$. \emph{Quijote} uses the TreePM \emph{Gadget-III} code and identifies halos using  a Friends-of-Friends \citep{1985ApJ...292..371D} halo-finding algorithm. Each simulation in the latin hypercube varies the cosmological parameters described in Tab.~\ref{tab:priors}, and the random phases of the initial conditions simultaneously. We randomly split the dataset 90\%/10\% into training and testing sets. For validation, we use $50$ simulations with varying random phases in the initial conditions and fixed cosmological parameters to their fiducial values.

Dark matter halo coordinates are represented as a 3-D point cloud, selecting the heaviest 5000 halos by halo mass. We chose to select halos by number density, as opposed to by chosing a minimum halo mass threshold, since in observations we only have access to the former. We also use the velocity and mass attributes from the halo catalogs for a subset of our experiments to demonstrate the ability of the model to reproduce correlations in a higher-dimensional feature space. Examples of samples from the test dataset are shown in the bottom row of Fig.~\ref{fig:boxes_pos}. 

\subsection{Diffusion-based generative modeling}
\label{sec:diffusion}

Diffusion models have emerged as state-of-the-art deep generative models in domains like computer vision, surpassing in flexibility and expressivity models like normalizing flows and variational autoencoders (VAEs). They admit several closely-related formulations. In one common framing \citep{ho2020denoising}, a neural network $\hat{{\epsilon}}_{{\varphi}}\left({z}_t, t\right)$ learns to iteratively `de-noise' a corrupted version $z_t$ of the data $x \equiv z_{t=0}$ from a timestep $t\in [0, T]$ by predicting either the additive noise $\epsilon$, the original data point $x$ directly, or some combination of the two \citep{salimans2022progressive}. New samples can then be generated by sampling Gaussian random noise $z_T$ and iteratively de-noising it from $t=T$ to $t=0$. A complementary framing \citep{song2020score} relies on having a neural network $\hat{{s}}_{{\varphi}}\left({z}_t, t\right)$ estimate the time-dependent gradient of the data distribution -- the so-called score function, $\nabla_{z_t} \log p(z_t)$.

The two formulations are closely related. Considering Gaussian noise addition with variance $\sigma_t^2$ as the forward process, $q(z_t \mid x) = \mathcal N(z_t; x, \sigma_t^2)$, the \emph{conditional} score can be analytically expressed as $\nabla_{z_t}\log q(z_t \mid x) = (x - z_t) / \sigma_t^2 = -\epsilon /\sigma_t$. Score- and noise-prediction are hence equivalent up to a timestep-dependent scaling. The intuition behind the relative negative sign is that, since the noise $\epsilon$ corrupts the data point, moving in its `opposite' direction will maximize the local (in time $t$) probability of moving towards the original data point. Hence, we refer to the noise- and score-prediction networks interchangeably.

\
\begin{deluxetable}{ccccc}[!ht]
\tablecaption{Definitions and ranges of the cosmological parameters of the \emph{Quijote} simulation suite.\label{tab:priors}}
\tablewidth{0pt}
\tablehead{
\colhead{} & \colhead{\textbf{Parameter}} & \colhead{\textbf{Interpretation}} & \colhead{\textbf{Range}} & \colhead{\textbf{Fiducial}}
}
\startdata
& $\Omega_{ m}$ & Matter density & [0.1, 0.5] & 0.3175 \\
& $\Omega_{b}$ & Baryon density & [0.03, 0.07] & 0.049 \\
& $h$ & Dimensionless Hubble constant & [0.5, 0.9] & 0.6711 \\
& $\sigma_8$ & Amplitude of matter fluctuations in $8\,h^{-1}{\rm Mpc}$ spheres & [0.6, 1.0] & 0.834 \\
& $n_s$ & Spectral index of the primordial power spectrum & [0.8, 1.2] & 0.9624 \\
\enddata
\end{deluxetable}

\subsection{Variational diffusion models}

Here, we use the \emph{variational diffusion model} formulation \citep{vahdat2021score,song2021maximum,kingma2023variational}, which frames the diffusion process as a hierarchical variational autoencoder (VAE) with a specific (Gaussian) functional form for the transition probability between latent variable hierarchies in the forward (noise-addition) process. Much as in a classical VAE \citep{kingma2013auto}, the evidence lower bound (ELBO) objective can be used as a variational lower bound on the log-likelihood $\log p(x)$. We give a high-level overview of the formalism here; see \citet{luo2022understanding,kingma2023variational} for further details. \\

\noindent
\textbf{The forward process:} 
The forward (noising) process is defined by the distribution $q\left(z_t \mid z_{t-1}\right)$, which also defined the \emph{noise schedule} of the diffusion model. This is a critical part of the model which can have a large impact on the final fidelity and learning dynamics of the model \citep{2023arXiv230110972C}. We take this to be a variance-preserving,
\begin{equation}
\label{eq:vpdiffusion}
    q\left(z_t \mid z_{t-1}\right)=\mathcal{N}\left(\sqrt{1-\beta_t} \cdot z_{t-1}, \beta_t\right)
\end{equation}
which corresponds to
\begin{gather}
    q\left(z_t \mid x\right) =\mathcal{N}\left(\sqrt{\bar{\alpha}_t} \cdot x, \sqrt{1-\bar{\alpha}_t}\right) \\
    \bar{\alpha}_t=\prod_{i=1}^t \alpha_i; \,\alpha_t\equiv 1-\beta_t.
\end{gather}
This is commonly referred to as the \emph{diffusion kernel} and can be used to conveniently predict a noised data sample at any timestep $t$ without going through intermediate times. We further define the signal-to-noise ratio or the mean-to-standard noise ratio, $\operatorname{SNR}(t)\equiv{\bar{\alpha}_t} / (1-\bar{\alpha}_t)$. \\

\noindent
\textbf{The variational objective:} 
For the diffusion objective, we use the efficient and numerically stable implementation of the variational evidence lower bound (ELBO) from \citet{kingma2023variational,2023arXiv230300848K}. The ELBO can be written as
\begin{equation}
\label{eq:elbo}
\log p(x) \geq \operatorname{ELBO}({x})=-\underbrace{\vphantom{\mathbb{E}_{q({z}_{t_1} \mid {x})}}\mathbb{E}_{q\left({z}_{T} \mid {x}\right)}\left[D_\mathrm{KL}\left(q\left({z}_T \mid {x}\right) \|\, p\left({z}_T\right)\right)\right]}_{\text {Prior matching }}
+\underbrace{\mathbb{E}_{q({z}_{t_1} \mid {x})}\left[\log p\left({x} \mid {z}_{t_1}\right)\right]}_{\text {Reconstruction }}
+\underbrace{\vphantom{\mathbb{E}_{q({z}_{t_1} \mid {x})}}\mathcal{L}_\mathrm{diffusion}({x})}_{\text {Forward-reverse consistency}}  
\end{equation}
where $z_T$ are the latent random variables at the last noising step, $z_{t_1}$ are the latent variables in the first noising step, and $q(z_t\mid x)$ are the (assumed Gaussian) variational posteriors on the noise addition. The prior-matching and reconstruction terms are exactly analogous to a classical VAE with a single bottleneck layer and contain no trainable parameters. The diffusion loss $\mathcal{L}_\mathrm{diffusion}({x})$ ensures consistency between the forward (noising) and reverse (de-noising) distribution at each step of the hierarchy \citep{luo2022understanding}, 
\begin{equation}
\label{eq:diff_loss}    
\mathcal{L}_\mathrm{diffusion}({x})=-\sum_{t=2}^T \mathbb{E}_{q\left(z_t \mid x\right)}\left[D_{\mathrm{KL}}\left(q\left(z_{t-1} \mid z_t, x\right) \|\,p_\varphi\left(z_{t-1} \mid z_t\right)\right)\right].
\end{equation}
The target denoising step $p_\varphi\left(z_{t-1} \mid z_t\right)$ is learned as an approximation the the ground truth $q\left(z_{t-1} \mid z_t, x\right)$, which corresponds to a local denoising of $z_{t}$ when we have access to the target image $x$. For Gaussian diffusion, it can be shown \citep{luo2022understanding} that the ground-truth denoising distribution can be written analytically as a Gaussian,
\begin{equation}
    q\left(z_{t-1} \mid z_t, x\right)=\mathcal{N}\left(z_{t-1} ; \mu_q\left(z_t, x\right), \sigma_q(t) \mathbb I \right),
\end{equation}
where we omit the functional forms of $\mu_q$ and $\sigma_q$ for brevity. If we also assume a Gaussian functional form for the learned transition distribution $p_\varphi(z_{t-1}\mid z_t)$, minimizing the KL-divergence in Eq. \eqref{eq:diff_loss} reduces to matching the means and variances of the two Gaussians. After some algebra, minimizing the target KL-divergence terms reduce to
\begin{equation}
    \label{eq:noise_matching}
    \argmin_\varphi D_{\mathrm{KL}}\left(q\left(z_{t-1} \mid z_t, x\right) \|\,p_\varphi\left(z_{t-1} \mid z_t\right)\right) = \argmin_\varphi \frac{1}{2 \sigma_q^2(t)} \frac{\left(1-\alpha_t\right)^2}{\left(1-\bar{\alpha}_t\right) \alpha_t}\left[\left\|\epsilon-\hat{\epsilon}_\varphi\left(z_t, t\right)\right\|^2_2\right]
\end{equation}
where $\hat{\epsilon}_\varphi\left(z_t, t\right)$ is the noise-prediction neural network that is optimized during training, parameterized by $\varphi$. In practice, the sum over timesteps in Eq. \eqref{eq:diff_loss} is computed as an expectation with appropriate scaling,
\begin{equation}
    \label{eq:diff_loss_exp}    
    \sum_{t=2}^T \mathbb{E}_{q\left(z_t \mid x\right)}\left[D_{\mathrm{KL}}\left(q\left(z_{t-1} \mid z_t, x\right) \|\,p_\varphi\left(z_{t-1} \mid z_t\right)\right)\right] \approx \frac{N_T}{2} \,\mathbb{E}_{t\sim \mathcal U\{1, T\},\, q\left(z_t \mid x\right)}\left[D_{\mathrm{KL}}\left(q\left(z_{t-1} \mid z_t, x\right) \|\,p_\varphi\left(z_{t-1} \mid z_t\right)\right)\right].
\end{equation}
This makes diffusion models especially efficient to train -- they don't require simulation of the entire trajectory back to the primal Gaussian at every training step unlike e.g, continuous-time normalizing flows \citep{2018arXiv181001367G}.

Finally, the pre-factor in Eq. \eqref{eq:noise_matching} can be elegantly written in terms of the time-dependent log-SNR, $\gamma(t)$, and the timestep discretization in Eq. \eqref{eq:diff_loss_exp} can be taken to the continuum limit, yielding the final diffusion loss
\begin{equation}
\label{eq:denoising}
\mathcal{L}_\mathrm{diffusion}({x})=\frac{1}{2} \mathbb{E}_{{\epsilon} \sim \mathcal{N}(0, \mathbb{I}),\,t \sim \mathcal U(0, T)}\left[\gamma_\eta'(t)\left\|\,{\epsilon}-\hat{{\epsilon}}_{{\varphi}}\left({z}_t, t\right)\right\|^2_2\right]
\end{equation}
where $\gamma_\eta'(t) \equiv \mathrm d \gamma_\eta / \mathrm d t$ is evaluated via automatic differentiation and the expectation over a standard Gaussian comes via the expectation over $q\left(z_t \mid x\right)$ in Eq. \eqref{eq:diff_loss}. Equation \eqref{eq:denoising} demonstrates that the variational likelihood objective is equivalent to the traditional denoising (noise-prediction) objective \citep{ho2020denoising}, with an appropriate weighting pre-factor \citep{2023arXiv230300848K}. 

The noise schedule $\gamma_\eta(t)$ is implicitly parameterized via the log-SNR, modeled as linearly increasing in time between learnable extremal values $\eta = \{\gamma_\mathrm{min}, \gamma_\mathrm{max}\}$ with $\gamma(t=T) = \gamma_\mathrm{min}$ and $\gamma(t=0) = \gamma_\mathrm{max}$. The noise-prediction neural network and noise schedule parameters $\{\varphi,\eta\}$ are hence simultaneously optimized towards the maximum-likelihood bounding objective. \\

\noindent
\textbf{Sample generation:}  With the learned transition step $p_\varphi\left(z_{t-1} \mid z_t; t, \theta\right)$ at hand, there are several ways of generating new samples. The simplest is perhaps through ancestral sampling: \emph{(1)} discretize the interval $[0, T]$ to a chosen number of timesteps, \emph{(2)} sample an initial random noise configuration $z_T\sim \mathcal N(0, \mathbb I)$, and \emph{(3)} run the reverse diffusion process by iteratively sampling $z_t\sim p_\varphi(z_{t-1}\mid z_t; t, \theta)$ until we arrive at a sample at $z_0 \equiv x \sim \hat p(x\mid\theta)$. We explicitly reinstate $\{t, \theta\}$-dependence here to emphasize that the transition distribution (through the noise-prediction network) is conditioned on the diffusion time $t$ and cosmological parameters $\theta = \{\Omega_m, \sigma_8\}$. An example of a sampled point cloud starting from a 3-D Gaussian sample is shown in Fig.~\ref{fig:samples}, along with the intermediate states. The sampled trajectories for a subset of particles are shown via the grey lines. See App.~\ref{app:pfode} for an alternative deterministic sampling method based on ODEs that produces smoother trajectories. In App.~\ref{app:pfode}, we also show how these smoother trajectories vary with a conditioning parameter, $\Omega_m$, to produce more or less clustered point clouds.\\

\noindent
\textbf{Likelihood evaluation:} The diffusion model is trained using a stochastic estimate of the variational maximum-likelihood objective, Eqs.~\eqref{eq:elbo} and \eqref{eq:diff_loss}. The same expression can be used to obtain an estimator of the conditional log-likelihood $\log \hat p(x\mid\theta)$, ensuring that the ELBO is evaluated a sufficient number of times to obtain a good estimate of the expectation value. In more detail, the interval $[0, T]$ is discretized into timesteps, and we iteratively draw $z_t \sim q(z_t\mid\, z_{t-1}, \theta)$ starting from $z_0 = x$, compute the diffusion loss terms in Eq. \eqref{eq:noise_matching}, which are summed up and added to the prior and recontruction terms in Eq. \eqref{eq:elbo}.

\subsection{The score/noise-prediction models}

The noise-prediction function $\hat{{\epsilon}}_{{\varphi}}\left({z}_t, t\right): \mathbb R^{N_\mathrm{gal}\times N_f} \rightarrow \mathbb R^{N_\mathrm{gal}\times N_f}$, where $N_\mathrm{gal}$ denotes the number of tracers and $N_f$ the number of features per tracer, is a crucial part of the diffusion model and must be chosen sensitive to the data modality and generating process. In our case, we model the distribution of tracers and their properties as a \emph{point cloud} i.e., a collection of coordinates (positions), optionally with attached attributes (e.g., velocities), $p\left(\{\vec r_i;[ \vec v_i,\ldots]\}_{i=1}^{N_\mathrm{gal}}\mid\,\theta\right)$. The number of features is either $N_f = 3$ when only modeling tracer coordinates, or $N_f=7$ when additionally modeling velocities and masses. 

The score model must be \emph{(1)} equivariant to permutations, \emph{(2)} able to process points of arbitrary cardinality, and \emph{(3)} able to effectively model the joint correlation structure of galaxy/halo properties. Variants of the closely related transformer and graph neural networks (GNNs) families satisfy these requirements; here, we show\textbf{} applications using both, described below. All models are implemented using the \package{Jax} \citep{jax2018github} framework. \\

\begin{figure*}
    \includegraphics[width=0.95\textwidth]{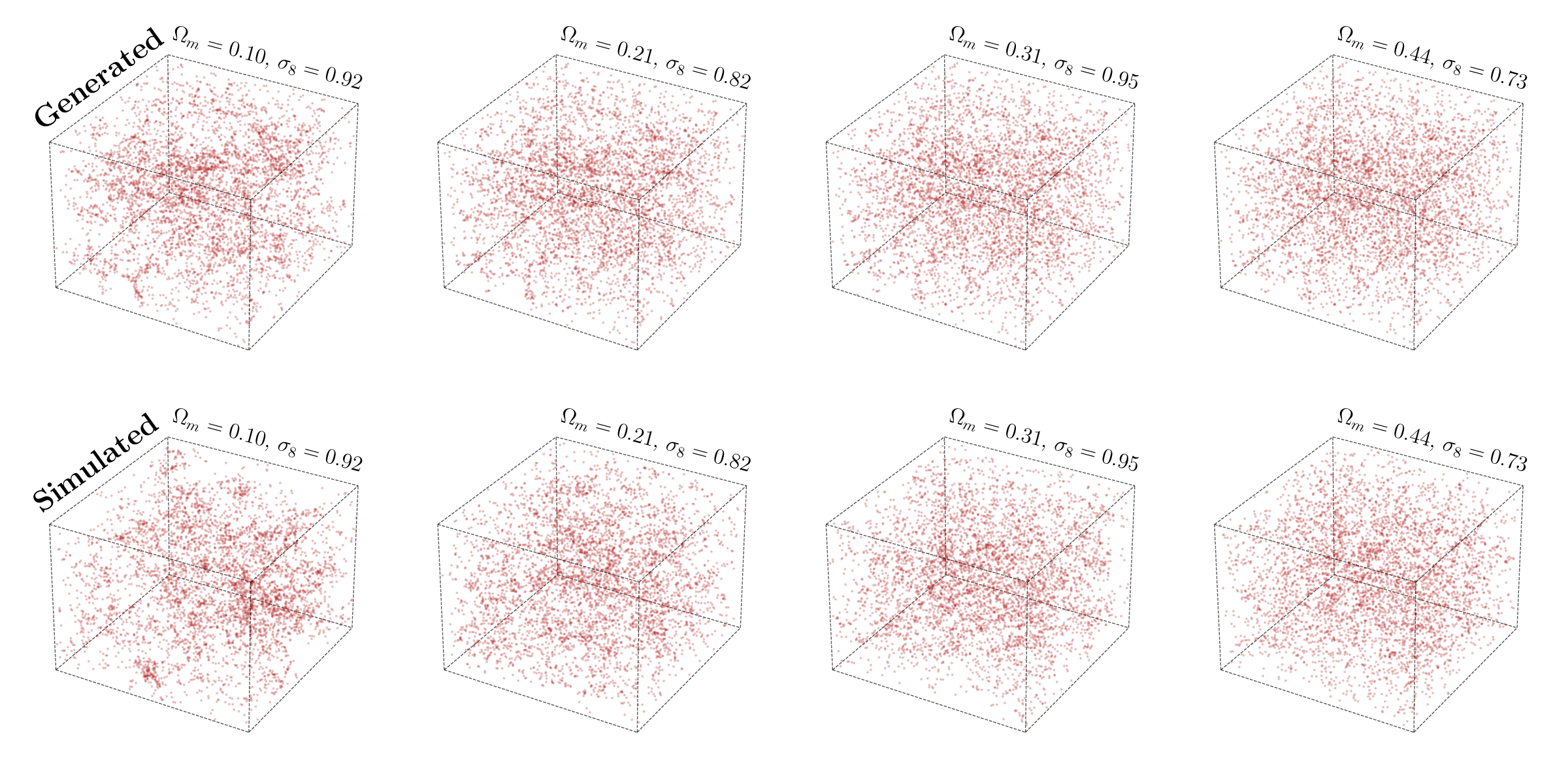}
\caption{Examples of point clouds generated from the trained position-only diffusion model (top row) and those from the test set (bottom row), with each column corresponding to the same set of cosmological parameters, indicated. The generated point clouds are drawn from the same random seeds. \nblink{04_viz_pos}}
\label{fig:boxes_pos}
\end{figure*}

\noindent
\textbf{Graph neural network model:} 
\label{sec:graph_score}
We use a variant of the graph-convolutional network from \citet{battaglia2018relational}. A local $k$-nearest neighbors graph with $k=20$ is constructed using the Euclidean distance between coordinates as the distance metric, accounting for periodic boundary conditions, at each time step in the diffusion process. The relative 3-D distances between input node coordinates are used as input graph edge features. The timestep $t$ is projected onto a 16-dimensional space via sinusoidal encodings, and the conditioning parameters $\theta = \{\Omega_m, \sigma_8\}$ are linearly projected also onto a 16-dimensional parameter space. They are concatenated to form the global conditioning vector $g^0$.
Both the input node features $z_t$ and edge features $z_{t,i}^{\mathrm{pos}} - z_{t,j}^{\mathrm{pos}}$ are initially projected into a 16-dimensional latent space via a 4-layer MLP (multi-layer perceptron; fully-connected neural network) with 128 hidden features and GELU activations. All MLPs utilized in the GNN have these same attributes.

Four message-passing rounds are performed, updating the edge attributes $e_{ij}$ at each round by passing a difference of the sender and receiver node attributes, edge attributes, as well as global parameters (a combination of timestep embedding and conditioning parameters $\{\Omega_m, \sigma_8\}$) through an MLP. For each node $h_i$, the neighboring edge attributes are aggregated, concatenated with the node and global attributes, and passed through another MLP to obtain the residual of the updated node features. Feature-wise layer normalization \citep{2016arXiv160706450L} is applied after each layer.

The graph convolutional layers are defined as 
\begin{gather}
    {e}^{l+1}_{ij} = \phi_e^l\left(\operatorname{Concat}[{h}^l_i - {h}^l_j, {e}^l_{ij}, g^0]\right) \label{eq:edge_update} \\
    {h}^{l+1}_{i} = {h}^{l}_{i} + \phi_h^l\left(\operatorname{Concat}\left[{h}^l_{i}, \sum_{j \in \mathcal{N}(i)_{/ i}} {e}^{l+1}_{ij},  g^0\right]\right)
\end{gather}
where the edge- and node-update neural networks $\phi_e^l$ and $\phi_h^l$ are both MLPs. \( \mathcal{N}(i) \) denotes the set of nodes connected to node \(i\) by an edge. Each edge and node update is additionally paramaterized by the global (diffusion time and conditioning) parameters $g^0$. Finally, the latent node features are projected back onto $N_f$ dimensions via an MLP. 

The GNN model integrates an attention mechanism to selectively emphasize relevant features in the graph when updating the edge features -- attention scores are computed for each edge and used to scale the edge features. First, for each edge \( (i, j) \) connecting nodes \(i\) and \(j\), an attention logit is calculated using an MLP \( \phi_a \),
\begin{equation}
    l_{ij}^l = \phi_a\left( \operatorname{Concat}[h_i^l -  h_j^l, e_{ij}^l, g^0] \right).
\end{equation}
These logits are then normalized across all neighboring edges via sofmax, ensuring that the attention scores for edges emanating from a single node sum to one. For a node \(i\), the attention weight \( \alpha_{ij} \) for each edge is given by $\alpha_{ij} = {\exp(l_{ij})}/\left({\sum_{k \in \mathcal{N}(i)} \exp(l_{ik})}\right)$.
The edge features are then scaled by these attention weights, $e_{ij}^{\prime} = e_{ij} \cdot \alpha_{ij}$,  resulting in attention-modified edge features which are used subsequently in Eq. \eqref{eq:edge_update}.  

The graph neural network was implemented using the \package{Jraph} \citep{jraph2020github} package and contains 637,373 trainable parameters.\\

\noindent
\textbf{Transformer model:}
\label{sec:transformer_score} The transformer \citep{2017arXiv170603762V,2023arXiv230410557T} is a sequence-to-sequence model that uses self-attention to process sequences of arbitrary length. We use a decoder-only transformer without positional encodings or causal masking, which makes the model permutation-equivariant and able to deal with set-valued data. The input coordinates $z_t$ are linearly projected onto an embedding space of dimension 256, then processed through 4 transformer layers each consisting of multi-head attention with 4 heads and a 2-layer multi-layer perceptrons (MLP) of hidden dimension 1024 with GELU activations. A `pre' layer norm configuration, where features are normalized each time the transformer residual stream is read from, was found to be crucial for training stability \citep{2020arXiv200204745X}. The final output is projected down to the dimensionality of the input attributes. In order to condition the score model on timestep $t$ as well as the parameters of interest $\theta = \{\Omega_m, \sigma_8\}$, a linear projection of the combined conditioning vector $g^0$ (described in the GNN model above) is added to the input embeddings. The transformer score model contains 4,776,281 trainable parameters.

Self-attention scales quadratically $\mathcal O(N_\mathrm{gal}^2)$ with the number of input points, in principle limiting the applicability of this architecture to larger point clouds. For set-valued data however, a specified number $N_\mathrm{ind}$ of representative `inducing' points can be learned also via attention -- essentially an on-the-fly learned clustering \citep{2018arXiv181000825L}. These are then used for computing the keys and values in the attention mechanism, with the input points projected onto queries, scaling the computation linearly $\mathcal O(N_\mathrm{gal} \cdot N_\mathrm{ind})$ with the cardinality of the point cloud. We implement and test induced attention in our code, achieving similar performance to full attention, but did not find it necessary for computational tractability with our 5000-cardinal point cloud.

\begin{figure*}
\label{fig:boxes_pos_vel}
\includegraphics[width=0.95\textwidth]{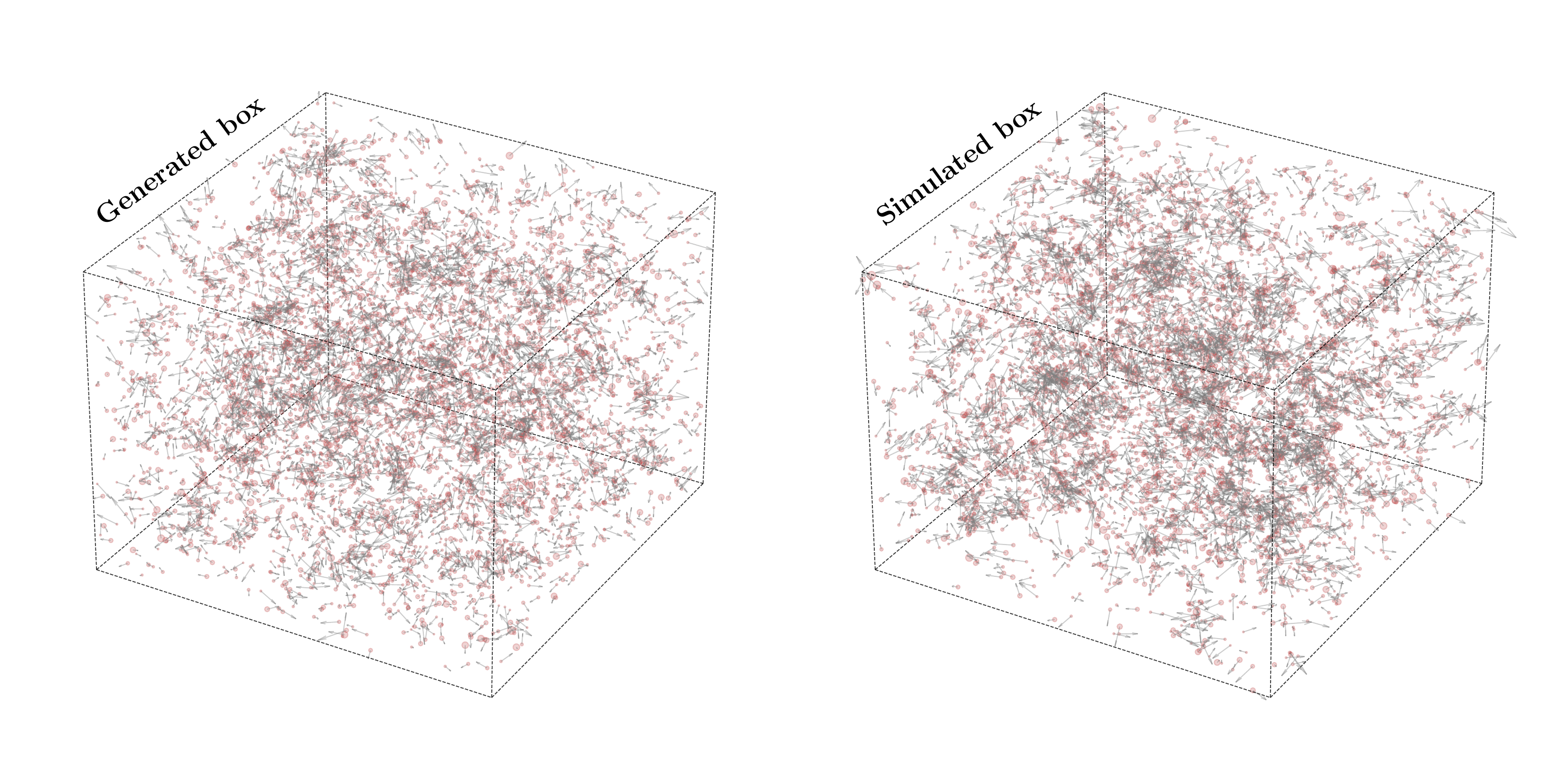}
\caption{Example of a point cloud generated from the diffusion model trained on positions, velocities, and masses (top row) and those from the test set (bottom row). Grey arrows correspond to the position and relative magnitude of velocities, and the the size of the individual points is proportional to the masses of the galaxies.  \nblink{03_viz_vel}}
\end{figure*}

\section{Results and discussion}
\label{sec:experiments}

\noindent
\textbf{Training:} The model is trained using the variational maximum-likelihood objective in Eq.~\ref{eq:elbo}. We run $300,000$ iterations of the AdamW \citep{DBLP:conf/iclr/LoshchilovH19,DBLP:journals/corr/KingmaB14} optimizer with peak learning rate $3\times10^{-4}$, $5000$ linear warmup steps, cosine annealing, and a batch size of 16. Boxes are randomly translated and rotated, sensitive to periodic boundary conditions, as a form of data augmentation. We select the checkpoint used downstream as the one with the smallest KL-divergence between two-point correlation functions of generated samples and those from the validation set. Further details are provided in App.~\ref{app:training}. We train models on either \emph{(1)}~halo positions only, or \emph{(2)}~halo positions, velocities, and masses. Training takes about 12 hours on 4 Nvidia A100 GPUs. \\

\noindent
\textbf{Conditional sampling:} Sampling a point cloud takes $\sim 5$ seconds on a single Nvidia A100 GPU using 1000 timesteps, scaling sub-linearly via vectorization when sampling batches. We show examples of position-only samples from our diffusion model, with a GNN backbone, in the top row of Fig.~\ref{fig:boxes_pos}, with the conditioning cosmological parameters $\{\Omega_m, \sigma_8\}$ annotated. Boxes from the test set corresponding to these parameters are shown in the bottom row. Generated boxes are drawn from the same random seed in order to emphasize the effect of parameter conditioning. We see clear signs of clustered structure, with the expected dependence as the cosmological parameters are varied. Although the overall clustering of dark matter particles increases with increasing $\Omega_m$, we here select the most massive 5000 dark matter haloes in each cosmology. In cosmologies with a lower value of $\Omega_m$ this selection will lead to smaller mass objects that tend to be near each other, as opposed to big clusters more separated in space for a larger $\Omega_m$ cosmology.

A sample from the diffusion model trained on positions, velocities, and masses with the transformer backbone is shown in Fig.~\ref{fig:boxes_pos_vel} (left) to be compared with a box from the test simulation suite with the same underlying cosmology (right). Velocity directions and magnitudes are indicated with grey attached arrows, and the size of the marker is proportional to the mass of the galaxy. Again, clear signs of clustered structure are visible.\\

\noindent
\textbf{Summary statistics validation:} We verify the quality of the trained generative model, including the dependence on cosmological parameters, by comparing the summary statistics obtained from the generated point clouds with those from a held out test set. 

\begin{figure*}[t!]
    \includegraphics[width=0.98\textwidth]{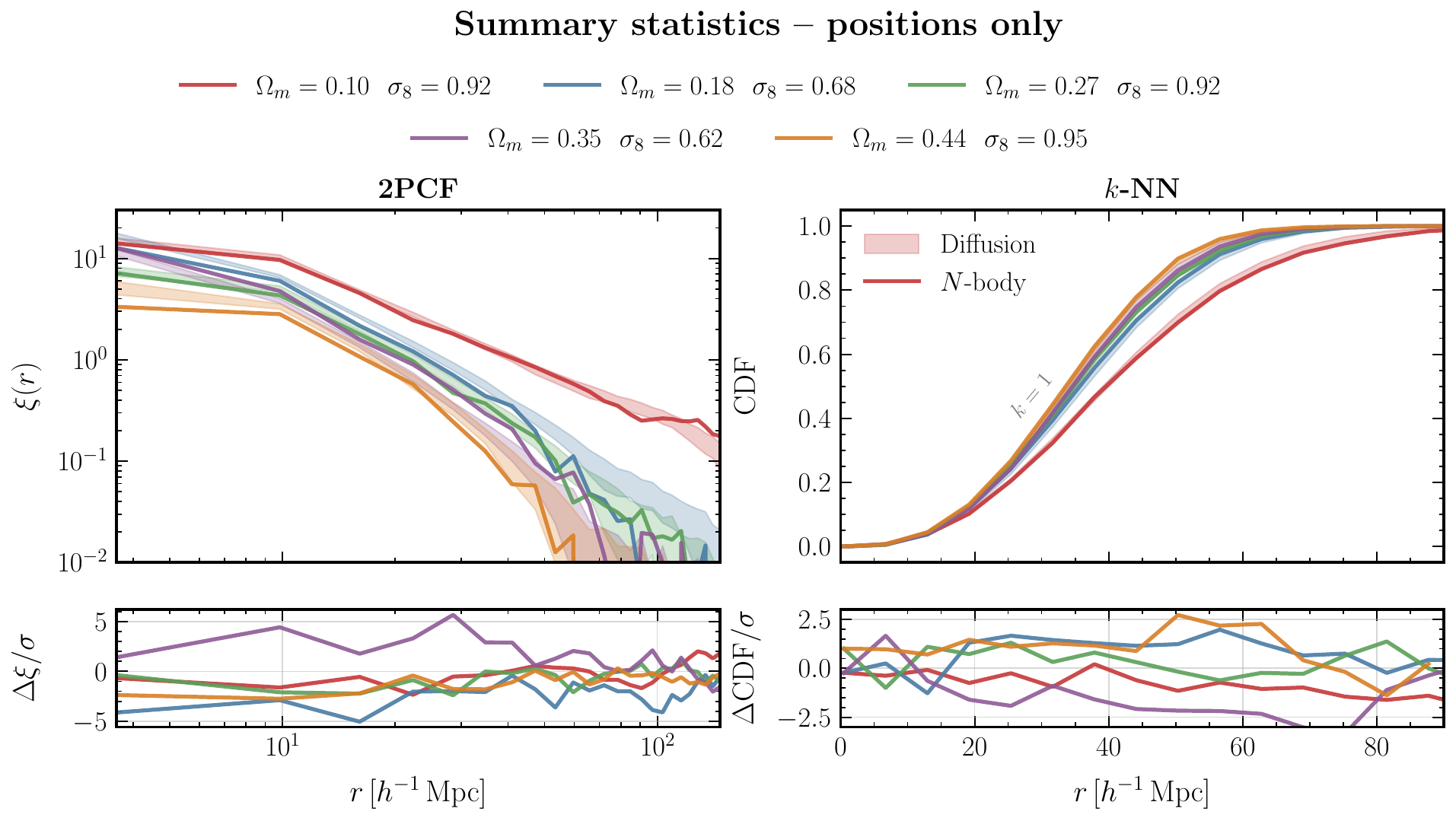}
\caption{Summary statistics of the samples generated by the diffusion model compared to those of the  $N$-body simulations for five equally spaced $\Omega_m$ values from the test set. For each cosmology, all summary statistics are computed for the same emulated point cloud. Lines are samples from the $N$-body simulations with different initial conditions, solid contours represent the mean and variance of $20$ samples from the diffusion model at that parameter value. On the left, we show the halo two-point correlation function. On the right, the cumulative density function for finding a first neighbour at a given distance from a random point in the simulation volume. Lower panels show the difference between the $N$-body and the mean of the diffusion samples, in units of the diffusion samples' standard deviation. \nblink{05_parameter_variations}}
    \label{fig:summaries}
\end{figure*}

In Fig.~\ref{fig:summaries}, we evaluate the positions only model trained with the  GNN score model described in Section~\ref{sec:graph_score}. We compare the parameter dependence of two widely used clustering statistics: the two-point correlation function (2PCF), and the cumulative distribution of $k$-nearest neighbors ($k$-NNs), evaluated at five different parameter values of the test set that have been chosen to span the $\Omega_m$ parameter space. We compare the mean and variance of $20$ diffusion samples to one sample from the $N$-body. Overall the diffusion model reproduces the trends of the $N$-body simulations, although due to the lack of varying seeds in the initial conditions for the $N$-body simulations at varying cosmological parameters we cannot provide a robust quantitative evaluation.  Note that the $N$-body samples also vary other cosmological parameters that are implicity marginalized over in the generated samples, namely, $\Omega_b$, $h$, and $n_s$. Hence we do not expect a perfect agreement between the two.

\begin{figure*}[t!]
    \includegraphics[width=0.98\textwidth]{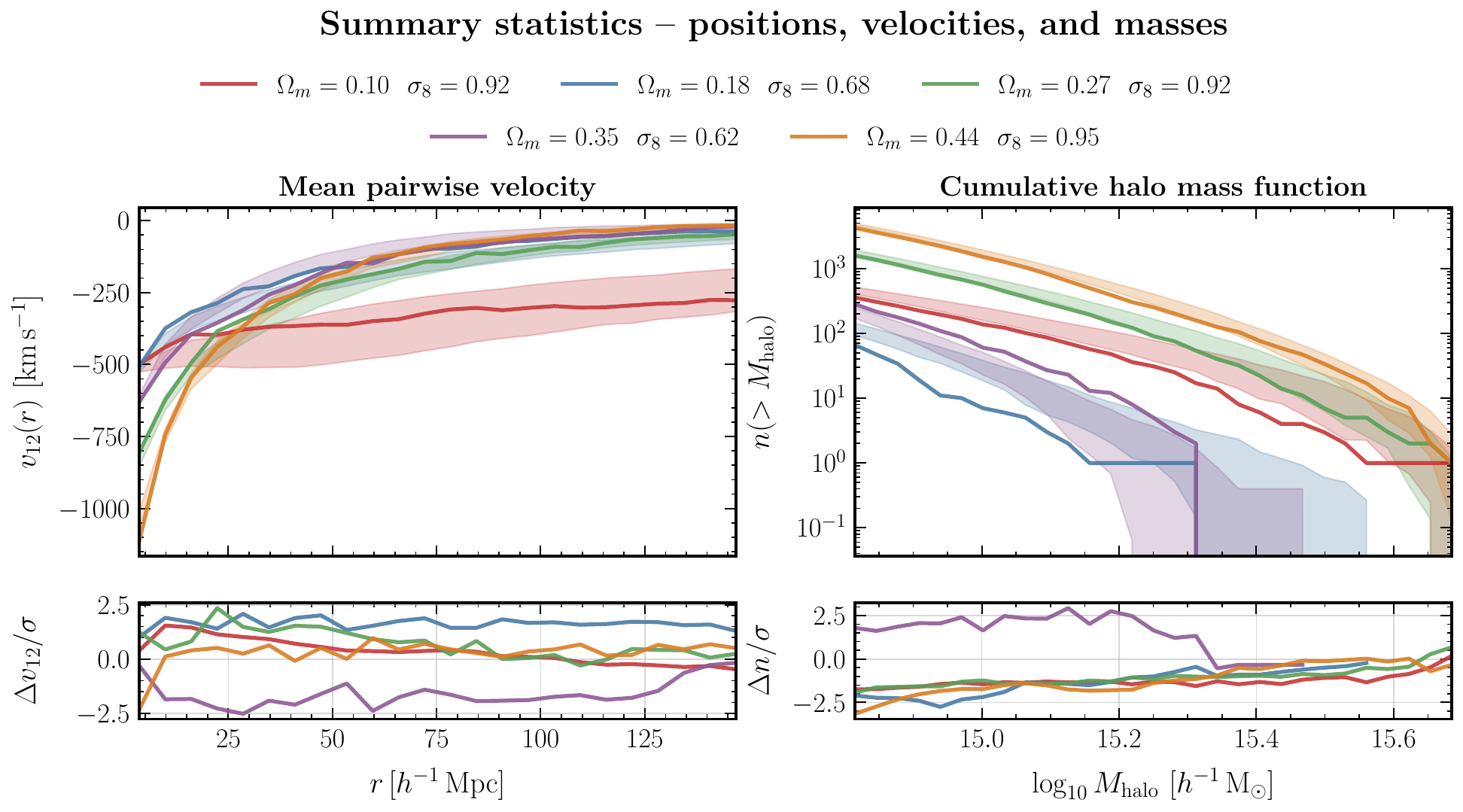}
\caption{Velocity (left) and mass (right) summary statistics of the samples generated by the diffusion model compared to those of the  $N$-body simulations for five equally spaced $\Omega_m$ values from the test set. On the left, we show the mean pairwise velocity as a function of pair separation. On the right, the cumulative halo mass function.  \nblink{05_parameter_variations}}
    \label{fig:velocity_summaries}
\end{figure*}

In Fig.~\ref{fig:velocity_summaries}, we show the corresponding evaluation for the model trained to reproduce the joint probability distribution of halo positions, masses and velocities, trained with the transformer score model described in Section~\ref{sec:transformer_score}. Here, we show the mean pairwise velocity distribution (left), to demonstrate that the model describes the joint distribution of velocities and positions faithfully, as well as the cumulative halo mass function (right). We find that the parameter dependence of the mean pairwise velocity can be reproduced over a wide range of scales, whereas the cumulative halo mass function seems to be slightly offset.

\begin{figure*}[htbp!]
    \includegraphics[width=0.98\textwidth]{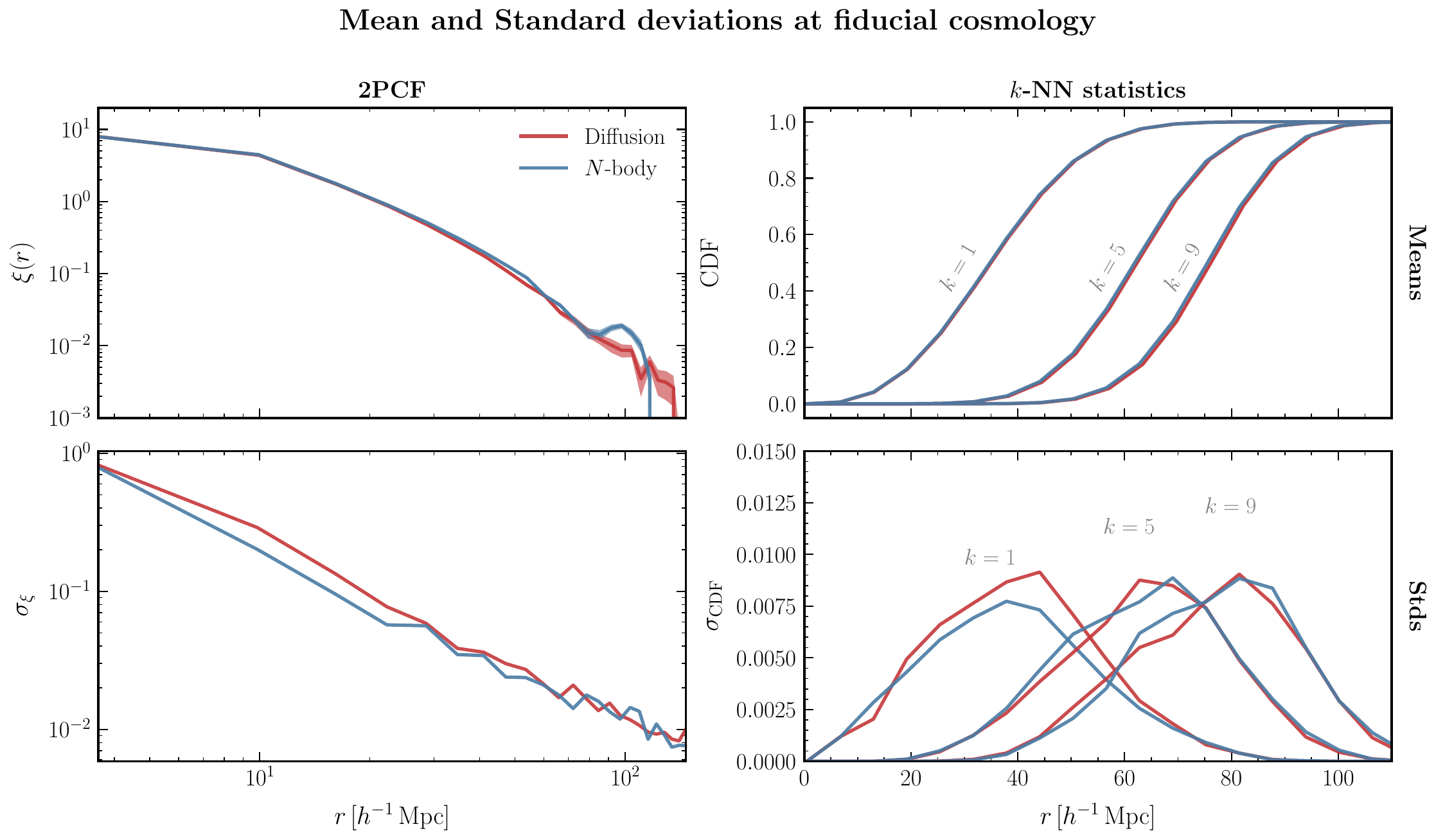}
    \caption{Mean and variances of the diffusion model and the $N$-body simulations for the two-point correlation function and the nearest neighbours statistics at the fiducial parameter values. Blue contours show the mean and variance of the Quijote simulations, whereas red contours show the mean and variance of the diffusion samples at the same parameter values. Upper pannels show a comparison of the mean, whereas lower pannels show differences in the standard deviation of the statistics as a function of scale. \nblink{06_fiducial_parameters}}
    \label{fig:summaries_fixed_params}
\end{figure*}

Finally, to assess the model's ability to model cosmic variance, we compare the mean and variance of $50$ diffusion samples' summary statistics to those of $50$ $N$-body simulations at the fiducial values, shown in Tab.~\ref{tab:priors}. In the first row of Fig.~\ref{fig:summaries_fixed_params}, we compare the mean of the 2PCF, and of the $k$-nearest neighbour statistics for different $k$ values, $k=1,5,9$. Although the model can reproduce the $k$-NN statistics well, it cannot accurately capture the behaviour of the 2PCF at the BAO scale ($\sim 120 \, h^{-1}\,\rm Mpc$). In the second row, we demonstrate that the model can indeed recover the standard deviation of the different summary statistics as a function of scale due to varying initial conditions for both the 2PCF and $k$-NN statistics. As already mentioned, the diffusion model is only conditioned on $\Omega_m$ and $\sigma_8$, and therefore is not expected to reproduce the fiducial distributions perfectly. \\

\begin{figure*}
        \includegraphics[width=0.95\textwidth]{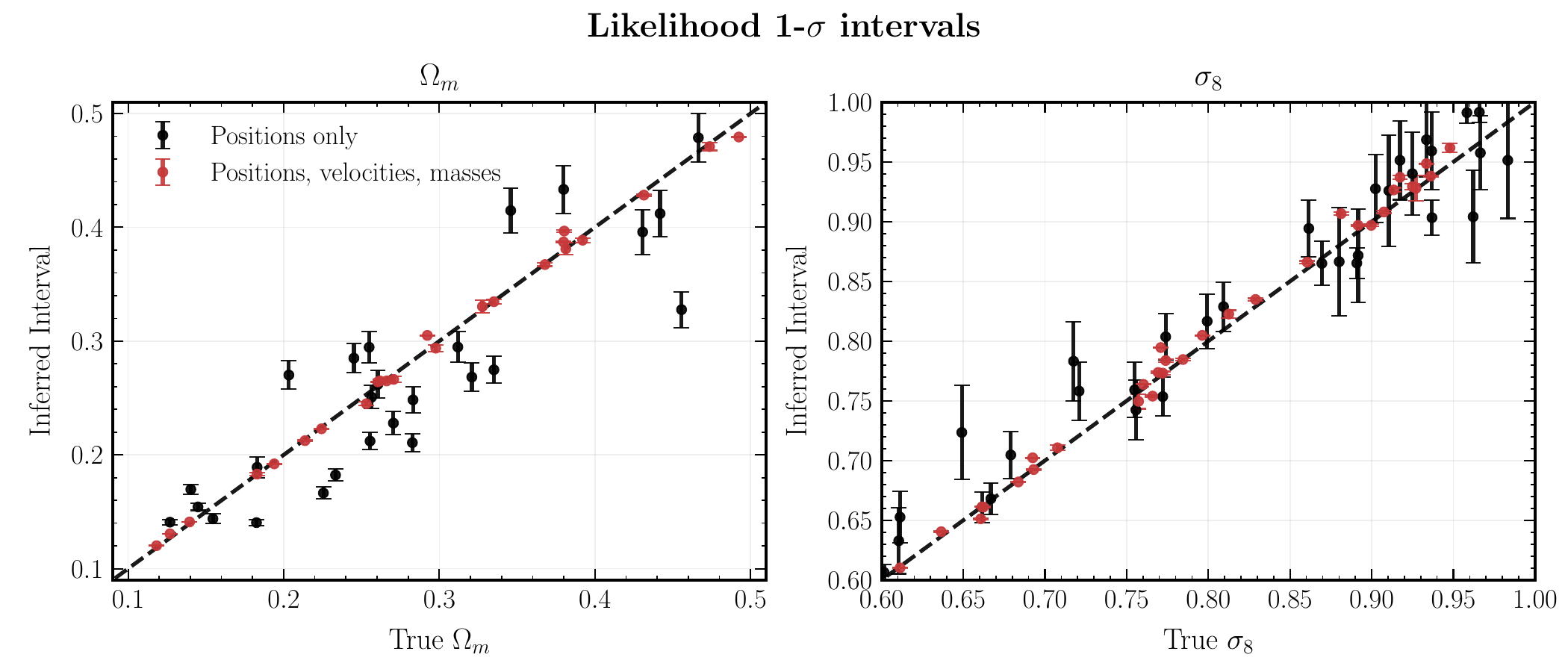}
    \caption{Intervals corresponding to 1-$\sigma$ containment from the likelihood profiles for $\Omega_m$ (left) and $\sigma_8$ (right), obtained by fixing one parameter at its true value and evaluating the likelihood over the other. Intervals for the position-only model (black data points) and model with positions, velocities, and masses (red data points) are shown. \nblink{01_likelihoods}}
    \label{fig:likelihoods}
\end{figure*}

\noindent
\textbf{Likelihood calculation:} Figure~\ref{fig:likelihoods} shows 1-$\sigma$ intervals on the parameters $\{\Omega_m, \sigma_8\}$ for samples from the held-out test set computed using the diffusion-backed approximate log-likelihood $\log\hat p(x\mid\theta)$ for the positions-only model (black data points) and model with all features (red data points). These are computed by varying the dependent parameter on a 1-D grid, while keeping the other parameter fixed at the ground truth value. The learned ELBO in Eq.~\eqref{eq:elbo} is evaluated 32 times with 50 discretization timesteps $\in [0,T=1]$ to obtain an estimate of the conditional log-likelihood, which takes $\sim 10$ seconds when vectorized over the number of evaluations. Additional results substantiating these choices are shown in App.~\ref{app:likelihood}

Although the points follow the expected trend qualitatively, it is clear that the learned likelihoods are not well calibrated. In particular, they are seen to be overconfident for $\Omega_m$. 

This is perhaps not surprising -- although we see from Fig.~\ref{fig:summaries} that the model has learned qualitatively good variation across cosmological parameters in the space of tested summaries, 1800 samples is likely too small a training dataset from which to robustly learn this parameter dependence at the field level, even with aggressive data augmentation. It can be seen from Fig.~\ref{fig:summaries_fixed_params} that the generated samples also struggle to reproduce the 2-point correlation function at larger scales, suggesting that the large-scale correlation structure, which influences in particular the $\Omega_m$ dependence, is not optimally captured.

Given this, we do not rigorously evaluate posteriors distributions on cosmological parameters or compare them to those obtained using 2-point correlation function summaries. We leave model improvements towards better-calibrated likelihoods to future work, some of which we discuss in the next section.

\section{Limitations and prospects}
\label{sec:prospects}

The present paper serves as a proof-of-principle exposition of some of the capabilities enabled by generative modeling in the context of galaxy surveys i.e., field level emulation and inference. Although we demonstrate the ability to emulate cosmological fields at the point cloud level, our GNN and transformer-backed models are not able to achieve well-calibrated likelihoods to a level that would be satisfactory for downstream applications, and also show discrepancies between the emulated and simulated point clouds on large scales. The availability of larger datasets for training is likely to partially alleviate this issue. We also outline promising directions on the methodological side for future study that could significantly improve the fidelity of our generative model and enable scaling to a larger number of points.

\begin{itemize}
    \item \textbf{Periodic boundary conditions:} The target point cloud data is confined to a box with periodic boundary conditions. Our model does not account for either the confinement to a box, or periodic boundary conditions at the level of the diffusion model (although note that the graph neighbor calculation does take into account distances across box boundaries). Existing methods used for generation of periodic configurations of materials \citep{2021arXiv211006197X,2023arXiv230702707L} could be leveraged in this direction. 
    \item \textbf{Physical symmetries:} Cosmological data typically encode a great deal of physical symmetry -- in our case, Euclidean symmetry associated with the freedom to choose an abitrary coordinate system. Our model is, on the other hand, manifestly coordinate dependent, relying on propagating absolute coordinates. Although we aim to partly mitigate this through data augmentation (i.e., train-time rotations, translations, and reflections), this is significantly less data-efficient than directly baking in these symmetries using symmetry-preserving neural networks \citep{2022arXiv220709453G,2021arXiv210209844G}, which have been shown to be provably more robust in other domains, e.g. the study of atomistic systems \citep{2022NatCo..13.2453B,2023NatCo..14..579M,2022arXiv220607697B}. Although translation- and rotation-invariant neural networks have previously been used within astrophysics for parameter prediction tasks via invariant feature propagation \citep{2022OJAp....5E..18M}, end-to-end \emph{equivariance} is expected to benefit our model more since it does not target prediction of globally-invariant properties.
    \item \textbf{Physically-motivated base distributions:} Our diffusion model relies on standard Gaussian diffusion, with the asymptotic latent distribution being a standard Gaussian $z_T \sim \mathcal N(0, \mathbb I)$. This means that the model has to denoise the standard Gaussian into a box. Recent classes of deep generative models e.g., stochastic interpolants and conditional flow matching \citep{2023arXiv230308797A,2023arXiv230200482T}, allow for the base distribution to be arbitrary and also implicitly defined through samples. In our case, using a physically-motivated initial distribution, e.g. particles in a box distributed according to a fiducial two-point correlation function, can be set up as a potentially easier learning problem.
    \item \textbf{Architecture expressivity:} Our fiducial score function is a simple message-passing GNN, which suffers from known lack of expressivity in particular when modeling long-range correlations. A common culprit is oversmoothing -- as the number of message-passing hops increases, the feature neighborhoods become increasingly similar across nodes, leading node features to collapse to similar values \citep{2023arXiv230310993K}. Indeed, our summary two-point correlation function does not faithfully capture correlation features on large spatial scales, such as the BAO peak (Fig.~\ref{fig:summaries_fixed_params}). The use of techniques to explicitly mitigate these issue \citep{2019arXiv190912223Z,2023arXiv230900367T} and enhance modeling of long-range correlation could therefore be helpful for generative modeling of cosmic fields as point clouds.
    \item \textbf{Scalability and hierarchical description of galaxy field:} Our diffusion model is used to represent the entire input point cloud field, which we choose to consist of 5000 particles. In practice, the field at large scales is expected to be highly linear and Gaussian, which calls for hierarchical methods that can generate fields described by nonlinear statistics on small scales while conforming to consistent linear descriptions on large scales. Designing the diffusion process in a lower-dimensional latent space via hierarchical down- and up-sampling~\citep{2017arXiv170602413Q}, also possible while preserving Euclidean symmetry~\citep{2023arXiv230504120F}, could be one avenue towards this.
\end{itemize}

\section{Conclusions}
\label{sec:conclusions}

We introduced a diffusion-based generative model that captures the complex, non-Gaussian statistics of the galaxy clustering field along with the underlying cosmology dependence. The model can be efficiently used for emulation of $N$-body simulations via sampling, $x \sim p(x\mid\theta)$, as well as evaluation of the conditional likelihood. While the model qualitatively reproduces essential summary statistics associated with the galaxy field, it can struggle to correctly model the point cloud's correlation structure in particular on larger spatial scales. We discuss the technical limitations of our model in this direcion, and avenues for further improvement.

The model presented in this work was trained on the dark matter halo distribution generated by $N$-body simulations. An application to upcoming galaxy clustering datasets, such as DESI, would require building a forward model for the survey that includes: \emph{(1)} a model of the galaxy-halo connection, \emph{(2)} observational effects, such as redshift space distortions and the Alcock-Paczynski effect, \emph{(3)} survey systematics, such as survey masks and fibre collisions. An example of such a forward model has been presented in \textsc{SimBIG} \citep{Hahn_2023}. A diffusion model for galaxy clustering trained on such a forward model could provide strong constraints on the standard $\Lambda$CDM cosmological model, as well as a means to test the robustness of its constraints through the analysis of posterior samples and likelihood estimates.

Code used for reproducing the results presented in this paper is available at \url{https://github.com/smsharma/point-cloud-galaxy-diffusion} \githubmaster. 

\begin{acknowledgments}
    This work is supported by the National Science Foundation under Cooperative Agreement PHY-2019786 (The NSF AI Institute for Artificial Intelligence and Fundamental Interactions, \url{http://iaifi.org/}). This material is based upon work supported by the U.S. Department of Energy, Office of Science, Office of High Energy Physics of U.S. Department of Energy under grant Contract Number  DE-SC0012567. The computations in this paper were run on the FASRC Cannon cluster supported by the FAS Division of Science Research Computing Group at Harvard University.
\end{acknowledgments}

\software{
    \package{Corrfunc} \citep{2020MNRAS.491.3022S}, \package{Diffrax} \citep{kidger2021on}, \package{Flax} \citep{flax2020github}, \package{GetDist} \citep{lewis2019getdist}, \package{Jax} \citep{jax2018github}, \package{Jraph} \citep{jraph2020github}, \package{Jupyter} \citep{Kluyver2016jupyter}, \package{Matplotlib} \citep{Hunter:2007}, \package{Numpy} \citep{harris2020array}, \package{NumPyro} \citep{phan2019composable}, \package{Optax} \citep{deepmind2020jax}, and \package{wandb} \citep{wandb}.
    }

\bibliography{set-diffuser}
\bibliographystyle{aasjournal-mod}

\appendix

\section{Alternative sampling method: probability flow ODE}
\label{app:pfode}

With the trained noise-prediction neural network $\hat{{\epsilon}}_{{\varphi}}\left({z}_t, t\right)$ at hand, several sampling techniques other than the ancestral sampling used in the main text are possible. Recall from Sec.~\ref{sec:diffusion} that the noise-prediction network is equal to the local conditional score of the data distribution, up to a sign and noise schedule-dependent scaling, $\nabla_{z_t} \log p(z_t) = \hat{{s}}_{{\varphi}}\left({z}_t, t\right) = -\hat{{\epsilon}}_{{\varphi}}\left({z}_t, t\right) / \sigma_t$.

In the continuous-time, SDE (stochastic differential equation) formulation \citep{song2020score}, taking the timestep discretization to the continuum limit $\delta_t \rightarrow 0$, the variance-preserving forward diffusion process in Eq.~\eqref{eq:vpdiffusion} can be written as 
\begin{align}
    z_t&=\sqrt{1-\beta_t \Delta_t} z_{t-1}+\sqrt{\beta_t \Delta_t} \epsilon \\
    &\approx z_{t-1}-\frac{\beta_t \Delta_t}{2} z_{t-1}+\sqrt{\beta_t \Delta_t} \epsilon
\end{align}
with $\epsilon\sim \mathcal N(0, \mathbb I)$. This is an update rule corresponding to the Euler-Murayama discretization of the SDE
\begin{equation}
\mathrm{d} z_t=-\frac{1}{2} \beta(t) z_t \mathrm{~d} t+\sqrt{\beta(t)} \mathrm{d} w_t
\end{equation}
where $w_t$ represents a Wiener process, also known as Brownian motion. This forward SDE has a corresponding \emph{reverse} SDE which gives the same marginal distribution $p(z_t)$ at all times,
\begin{equation}
    \mathrm{d} z_t=\left[-\frac{1}{2} \beta(t) z_t-\beta(t) \nabla_{z_t} \log p(z_t)\right] \mathrm{d} t+\sqrt{\beta(t)} \mathrm{d} w_t.
\end{equation}
This SDE can be solved with with any solver, and in the case presented via simple discretization (Euler-Murayama method).

Remarkably, there exists a deterministic reversible process, an ODE (ordinary differential equation), whose trajectories share the same marginal densities $p(z_t)$ as this reverse SDE~\citep{song2020score,song2021maximum},
\begin{equation}
    \mathrm{d} z_t=-\frac{1}{2} \beta(t) \left[ z_t +  \nabla_{z_t} \log p(z_t)\right] \mathrm{d} t
\end{equation}
called the probability flow ODE and which is an instance of a continuous-time normalizing flow \citep{2018arXiv181001367G}.

\begin{figure}[!tb]
    \centering
    \includegraphics[page=2, width=0.95\textwidth]{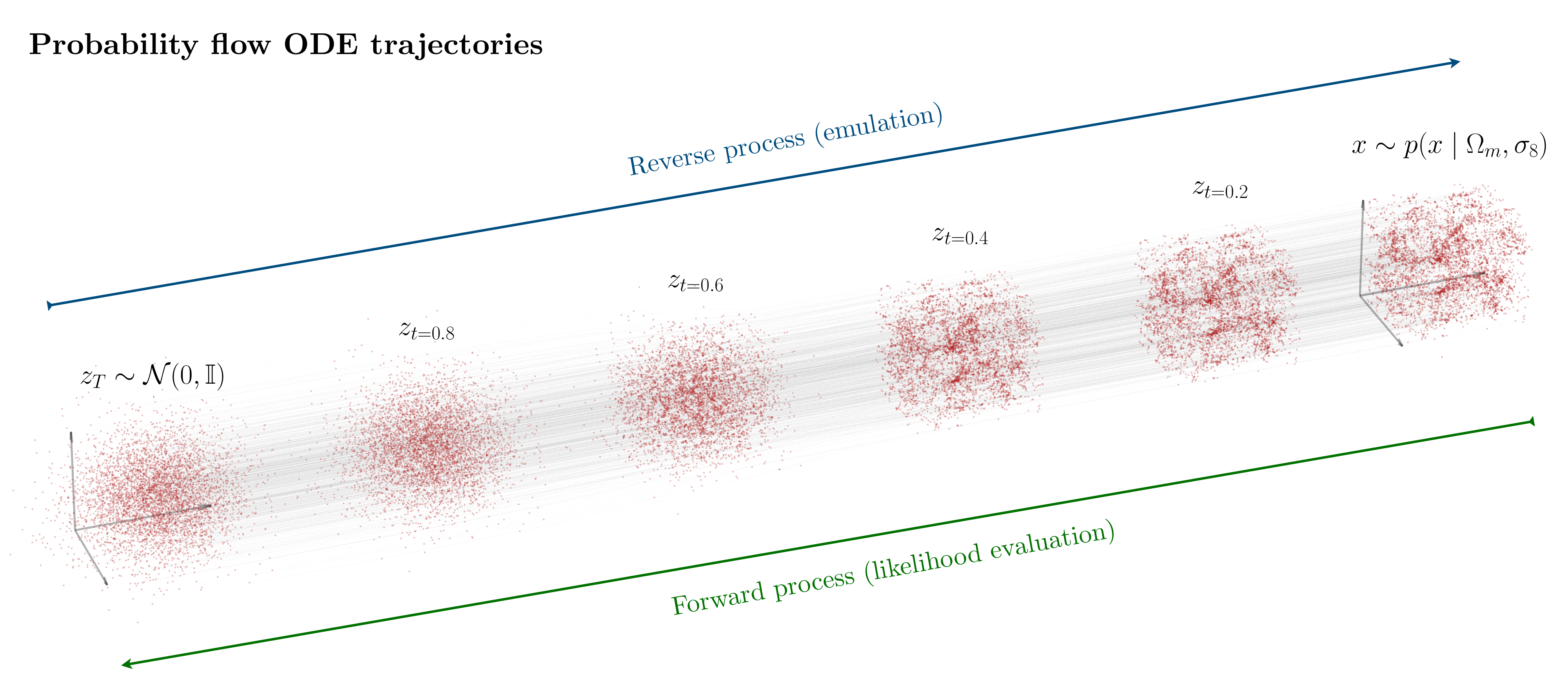}
    \caption{A sample obtained by solving the probability flow ODE associated with the diffusion SDE. Smooth trajectories from the primal Gaussian (left) to the sampled halo distribution (right) can be seen, in contrast with the stochastic trajectories in Fig.~\ref{fig:samples}.  \nblink{04_viz_pos}}
    \label{fig:pfode}
\end{figure}

This highlights the fact that diffusion models essentially efficiently train a continuous normalizing flow without requiring simulation the entire forward trajectory. This ODE can be solved using any ODE solver to obtain samples from the data distribution; we show in Fig.~\ref{fig:pfode} a sample and corresponding trajectories for a subset of particles obtained using Heun's second order method via \package{Diffrax} \citep{kidger2021on}. The trajectories can be seen to be smooth, in contrast with those in Fig.~\ref{fig:samples}. 

\begin{figure}[!ht]
    \centering
    \includegraphics[page=3, width=0.95\textwidth]{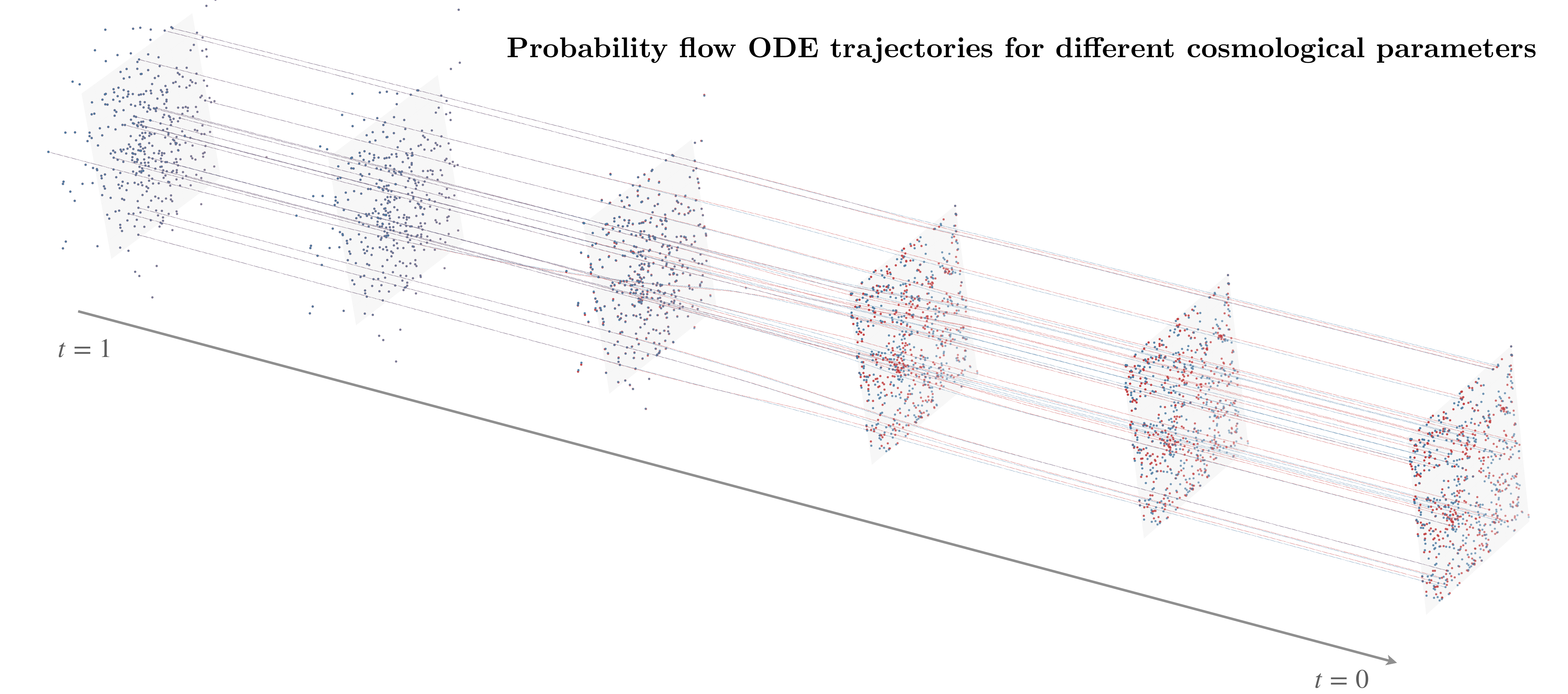}
    \caption{Probability flow ODE trajectories starting from the same initial Gaussian distribution, conditioned on different cosmological parameters with $\Omega_m = 0.13$ (red) and $\Omega_m = 0.47$ (blue), keeping fixed $\sigma_8=0.8$. 2-D spatial projections for 400 particles are shown, with trajectories illustrated for a subset of these.  \nblink{04_viz_pos}}
    \label{fig:pfode_cosmo}
\end{figure}

Sampling via solving the probability flow ODE can be used to visualize the dependence of the sampling trajectory on cosmological parameters, unencumbered by the stochastic component. This is shown in Fig.~\ref{fig:pfode_cosmo}, which illustrates 2-D projected slices of 400 generated particle coordinates across timesteps along with the sampled trajectories for a subset of the particles. Starting from the same initial Gaussian distribution, we condition on different cosmological parameters with $\Omega_m = 0.13$ (red) and $\Omega_m = 0.47$ (blue), keeping fixed $\sigma_8=0.8$. Diverging trajectories across diffusion time can be seen, highlighting the parameter-dependence of the score model.

\section{Additional training details}
\label{app:training}

\begin{figure}[!ht]
    \centering
    \includegraphics[width=0.95\textwidth]{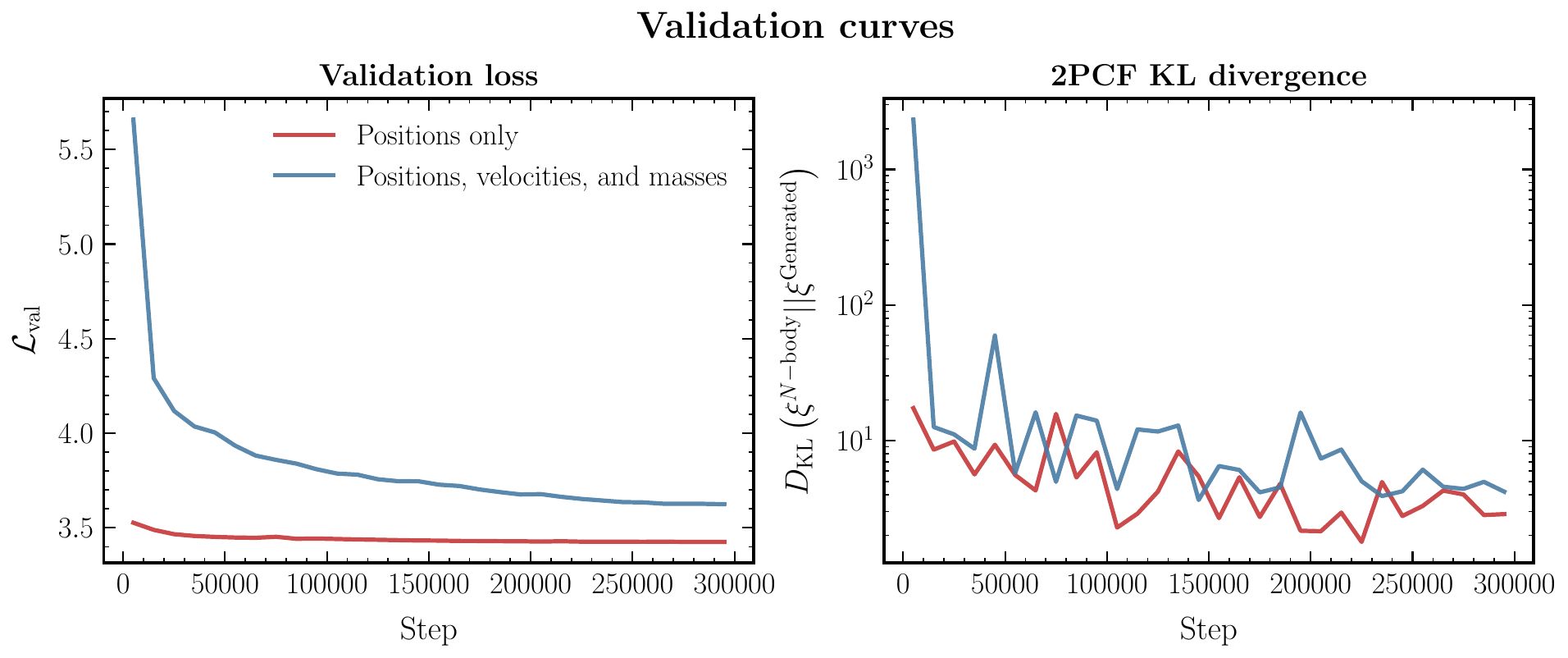}
    \caption{Validation loss (left) and KL-divergence between the true two-point correlation functions and the generated ones (right) as a function of training steps. We show both a graph neural network model trained on halo positions only (red), and a transformer model trained on positions, velocities, and masses (blue). \nblink{07_loss_KL}}
    \label{fig:loss_kl}
\end{figure}

In Fig.~\ref{fig:loss_kl}, we show the validation loss curve, together with the KL-divergence between the $N$-body two-point correlation functions on small scales ($r < 55 \,h^{-1}\,{\rm Mpc}$) and the generated ones. The KL-divergence is computed assuming that both distributions are Gaussian. We show that a small decrease in the loss value produces a sharp decrease in the KL-divergence and therefore a large improvement in the quality of the generated samples. The checkpoint used downstream for each model is chosen as the one with the smallest KL-divergence in the validation set.

\section{Log-likelihood evaluation}
\label{app:likelihood}

The diffusion model is trained by maximizing a stochastic estimate of a variational lower bound on the log-likelihood, the ELBO in Eq.~\ref{eq:elbo}. The same expression can be used to compute an estimate of the (conditional) log-likelihood, which we show as delta likelihood profiles in Fig.~\ref{fig:likelihoods} when averaged over 32 (dashed red) and 64 (solid red) evaluations for $\Omega_m$ (left) and $\sigma_8$ (right). 50 discretization steps are used; we found quantitatively similar results for the mean profiles using a larger number of steps. Different evaluations are shown as grey lines, demonstrating that the variance of the approximate likelihood with respect to the random seed is high relative to the $1\sigma$ interval. It can be seen that the relative log-likelihood has converged with 32 evaluations, which we use to show the likelihood profile results in Fig.~\ref{fig:likelihoods}.

Interestingly, variation on the \emph{absolute} log-likelihood estimate was observed to significantly larger, $\mathcal O(100)$, compared to the $\mathcal O(1)$ variation on the relative conditional log-likelihood shown. An estimate of the raw log-likelihood would therefore require a larger number of evaluations in this case.

\begin{figure}[!ht]
    \centering
    \includegraphics[width=0.95\textwidth]{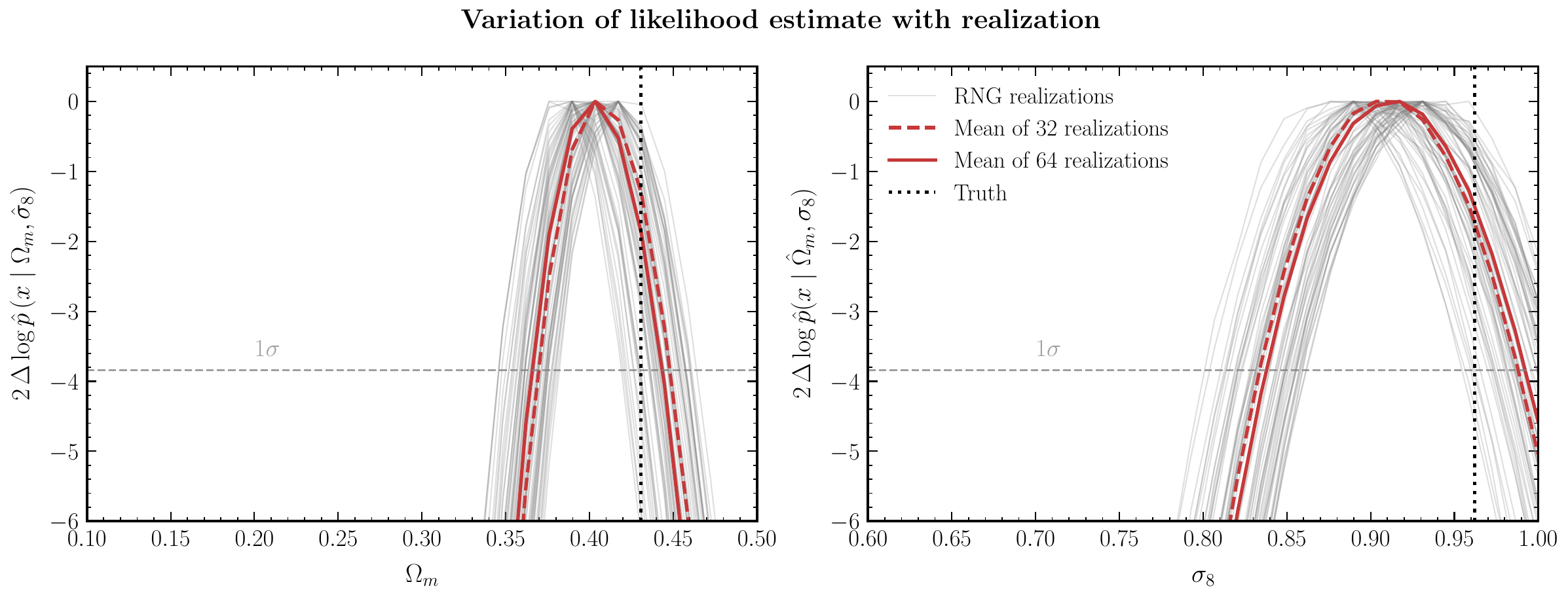}
    \caption{For a particular validation sample, twice the log-likelihood profiles relative to their maximum value, for $\Omega_m$ (left) and $\sigma_8$ (right). Profiles for different evaluations (grey) as well as averaged over 32 (dashed red) and 64 (solid red) evaluations are shown. Averaging over 32 realizations is seen to give a converged estimate of the conditional likelihood.  \nblink{02_likelihood_variation}}
    \label{fig:ll_realiz}
\end{figure}

\end{document}